\documentclass[preprint]{aastex}

\usepackage{epstopdf}
\usepackage{graphics}
\usepackage{lscape}
\usepackage{psfig}

\def\grb{GRB\,130603B}

\newcommand{\gp}{\mbox{$g^{\prime}$}}
\newcommand{\rp}{\mbox{$r^{\prime}$}}
\newcommand{\ip}{\mbox{$i^{\prime}$}}
\newcommand{\zp}{\mbox{$z^{\prime}$}}
\def\xray{X-ray}
\def\swift{\emph{Swift}}
\def\lya{\hbox{{\rm Ly}\kern 0.1em$\alpha$ }}
\def\NII{\hbox{{\rm N}\kern 0.1em{\sc ii}}}
\def\OII{\hbox{[{\rm O}\kern 0.1em{\sc ii}]}}
\def\OIII{\hbox{[{\rm O}\kern 0.1em{\sc iii}]}}
\def\Hb{\hbox{\rm H$\beta$}}
\def\Ha{\hbox{\rm H$\alpha$}}
\def\FeII{\hbox{{\rm Fe~}\kern 0.1em{\sc ii}}}
\def\MgII{\hbox{{\rm Mg~}\kern 0.1em{\sc ii}}}
\def\CaII{\ion{Ca}{2}}

\newcommand{\sfr}{$1.84\, {\rm M}_{\odot} \,{\rm yr}^{-1}$\,}
\newcommand{\zgrb}{$z = 0.3568$}

\begin{document}

\title{Gemini Spectroscopy of the Short-Hard Gamma-ray Burst 
GRB 130603B Afterglow and Host Galaxy}


\author{
	A.~Cucchiara\altaffilmark{1}, 
	J.~X.Prochaska\altaffilmark{1},
	D.~Perley\altaffilmark{2,3},
	S.~B. Cenko\altaffilmark{4},
	J.~Werk\altaffilmark{1},
	A.~Cardwell\altaffilmark{5},
	J. Turner\altaffilmark{5},
	Y.~Cao\altaffilmark{2},
	J.~S. Bloom\altaffilmark{6},
	B.~E. Cobb\altaffilmark{7}
	}
	\email{acucchia@ucolick.org}

\altaffiltext{1}{Department of Astronomy and Astrophysics, UCO/Lick Observatory, University of California, 1156 High Street, Santa Cruz, CA 95064, USA}
\altaffiltext{2}{Department of Astronomy, California Institute of Technology, MC 249-17, 1200 East California Blvd, Pasadena CA 91125, USA}
\altaffiltext{3}{Hubble Fellow}
\altaffiltext{4}{Astrophysics Science Division, NASA Goddard Space Flight Center, Greenbelt, MD, USA} 
\altaffiltext{5}{Gemini South Observatory, AURA, Casilla 603, La Serena, Chile} 
\altaffiltext{6}{Department of Astronomy, University of California, Berkeley, CA 94720-3411, USA} 
\altaffiltext{7}{The George Washington University, Washington, D.C., USA} 

\begin{abstract}
We present early optical photometry and spectroscopy of the afterglow
and host galaxy of the bright short-duration gamma-ray burst \grb\ discovered 
by the \swift\ satellite.
Using our Target of Opportunity program on the Gemini South telescope,
our prompt optical spectra reveal a strong trace from the afterglow
superimposed on continuum and emission lines from the $z = 0.3568 \pm
0.0005$ host galaxy.  The combination of a relatively bright optical
afterglow ($r^{\prime} = 21.52$ at $\Delta t = 8.4$\,hr), together
with an observed offset of 0\farcs9 from the host nucleus (4.8\,kpc
projected distance at \zgrb), allow us to extract a relatively
clean spectrum dominated by afterglow light .

Furthermore, the spatially resolved spectrum allows us to constrain
the properties of the explosion site directly, and compare these with
the host galaxy nucleus, as well as other short-duration GRB host
galaxies.  We find that while the host is a relatively luminous ($L
\approx 0.8 L^{*}_{B}$), star-forming (SFR = \sfr) galaxy with almost solar metallicity,
the spectrum of the afterglow exhibits weak \CaII\ absorption features but 
negligible emission features.  The explosion site therefore lacks evidence of
recent star formation, consistent with the relatively long delay time
distribution expected in a compact binary merger scenario. 
The star formation rate (both in an absolute sense and normalized to the
luminosity) and metallicity of the host are both consistent with the
known sample of short-duration GRB hosts and with 
recent results which suggest GRB\,130603B emission to be the product of  
 the decay of radioactive species produced during the merging process of a NS-NS binary (``kilonova'').  
 Ultimately, the discovery of more events similar to \grb\ and their rapid follow-up from 8-m class telescopes
   will open new opportunities for our understanding of the final stages of compact-objects binary systems
   and provide crucial information (redshift, metallicity and chemical content of their explosion site) to characterize 
   the environment of  one of the most promising gravitational wave sources.

\end{abstract}

\keywords{gamma-ray burst: individual (GRB 130603B)  -  techniques:
  imaging -  techniques: spectroscopy }


\section{Introduction }
\label{sec:intro}
Short gamma-ray Bursts (SGRBs) are historically identified based on the duration
of their gamma-ray emission ($T_{90} \lesssim 2$\,s\footnote{Time over which a burst emits from 5\% of its total measured counts to 95\%}) and their hard 
spectrum \citep[e.g.][]{Mazets:1981fk,Norris:1984uq,kmf+93}.
Recently, based on a statistical approach, several attempts have been made to improve this classification
\citep{Ripa:2009ys,Veres:2010zr,Bromberg:2013ve}.
Short-GRBs also differ from the ``long'' GRBs class for their redshift distribution
and, likely, their progenitors \citep[see, for example,][ and references therein]{Li:1998qf,OShaughnessy:2008kx,Zhang:2009vn,Kann:2011qf}.
 
The optical afterglows of SGRBs are on average two orders of 
magnitude less optically luminous than their long duration counterparts \citep{Kann:2011qf}, 
making broadband follow-up, and optical spectroscopy in particular, quite challenging.
Nevertheless, the {\it Swift} satellite \citep{Gehrels:2004fj}
has enabled the localization
of a modest sample in X-rays and a smaller set have been detected 
in optical/near-IR passbands
\citep[e.g.][]{Fox:2005dq,Bloom:2006pd,Hjorth:2005fk,Berger:2007uq,Nakar:2007vn,Fong:2013kx}.

In contrast, long GRBs  ($T_{90} \gtrsim 2$\,s) present brighter
afterglows allowing accurate localization and spectroscopic follow-up
hours after the events occur.
Robotic facilities and Target of Opportunity (ToO)
programs have provided a plethora of photometric and spectroscopic data 
 in support of theoretical models of long GRB progenitors and the 
host galaxies they live in. These data first
established conclusively the extragalactic nature of the events
\citep{mdk+97} and, eventually, analysis of the \lya\ forest for
high-$z$ events \citep[e.g.][]{Salvaterra:2009dz,Tanvir:2009fv,Cucchiara:2011uq}
and fine-structure transitions at lower redshift provided unambiguous physical associations to their hosts.
Optically bright long GRBs seem exclusively hosted by star-forming galaxies with high specific 
star-formation and sub-L$^*$ luminosities, indicating massive stars as 
likely progenitors of long GRBs \citep[see ][ and reference therein]{Levesque:2013fu}, 
while ``dark'' GRBs \citep{Jakobsson:2004uq} seem to be harbored, on average, 
in more massive and highly star-forming galaxies   
 \citep[$3\times 10^{10} $\,M$_{\odot}$ at $z\approx2$,][]{Perley:2013fk}.


Thanks to a concerted community effort, of the $\sim70$ short GRBs
detected by \swift\,  $\sim 1/3$ have been localized 
to within a few arcseconds accuracy, and, with a  {\it a posteriori} probabilistic 
arguments, have been securely associated with nearby galaxies \citep[see, e.g.,][]{Fong:2013kx,Fong:2013kz}.
The number of these events, however, is still in the few dozens \citep{Fong:2013kx}.
The few well-observed SGRBs have been associated primarily to a population 
of galaxies very similar to field galaxies at similar redshifts with moderate to negligible SFR, lending credence to the idea that at least some SGRBs explode with delay times 
$\gtrsim1$ Gyr consistent with the progenitor model of compact mergers 
 \citep[e.g.,  neutron star binaries,
or neutron star-black hole][]{Lee:2005qa, Hjorth:2005fk, Lee:2010ys}.
Recently, using deep observations
from space and from the ground it has been possible to quantify the relative fraction
short-duration GRB hosts: $20-40\%$ early- and $60-80\%$ late-type galaxies \citep{Berger:2010cr,Fong:2013kx}.

From the afterglow perspective, there has yet to be a {\it bona fide} short 
duration GRB afterglow for which 
we have measured a redshift from absorption features
 in the optical spectrum. Among these it is important to note the long GRB\,090426, for which the duration of the prompt emission ($t_{90}$ = 1.2\,s) and the properties of the host made its classification uncertain \citep[][]{Antonelli:2009ly,Levesque:2010kl}.
Another debated short GRB for which an afterglow+host galaxy spectrum has been
obtained is GRB 051221A \citep{Soderberg:2006kl}: also in this case the non-collapsar
nature of this event has been recently questioned based on probabilistic arguments and 
an accurate analysis of the instrumental biases which may lead one to mistakenly associate 
collapsar events like this one to short-GRBs \citep{Bromberg:2013ve}.
Finally, GRB\,100816A, despite having $t_{90} \approx 2$\,s, has been 
associated with the SGRB class based on lag analysis \citep{Norris:2010bh} and
its afterglow (or a combination of afterglow and host) has been spectroscopically observed 
\citep{Tanvir:2010qf,Gorosabel:2010dq}. 
The lack of a large sample of SGRB afterglow spectra has made 
it impossible to conduct analogous studies of the host galaxy
ISM and circumburst environment, as are routinely achieved for
long-duration events.

Finally,  the lower rate of  space-based localization (compared to
the long GRB class) and their faintness 
demand a very rapid response by large area
telescopes to reach the quickly fading afterglows and obtain 
similar high-quality data for a larger sample of SGRBs. 
Such a collection will serve two main goals: 1) provide  
unambiguously, based, e.g., on absorption-line diagnostics like fine-structure 
transition, the redshift of
these events and their association with 
an underlying host galaxy; 2) directly allow the characterization of the interstellar and
circumstellar environment from 200-300pc up to the host halo \citep{Chen:2005vn,Prochaska:2006fk,Berger:2006uq,Vreeswijk:2013kx} and, 
then, provide clues about the progenitor itself.

In this letter we present our prompt optical spectroscopic and photometric
observations of the short \grb, obtained at the Gemini South 
telescope. Based on the optical and near infrared emission  
at late time ($\sim 9$ days post-burst) derived by deep \emph{HST}
observations which allowed to clearly resolve the GRB-host complex, it has been proposed that this event might resemble
the expected emission due to the decay of radioactive species produced and initially ejected
during the merging process of a neutron star's binary system, referred to as a ``kilonova''
\citep{Li:1998qf,Metzger:2012ve,Berger:2013ij,Tanvir:2013hc}.

Thanks to our ToO program, we  observed this event within the first day,
when the afterglow dominated the host flux despite their small projected separation.
This provides constraints on the event redshift and the properties of the GRB 
explosion site, in particular, in comparison to the overall SGRB host galaxy population.

The paper is divided as follows: in \S\ref{sec:data} we present our
observing campaign; in Sec. \S\ref{sec:analys}
we present our spectral analysis, and
in \S\ref{sec:results} we discuss our results and compare them with previous studies on SGRBs and 
their host galaxies. Finally, in \S\ref{sec:conclusions} we present
some implications on the nature of \grb\ and the possibilities offered by rapid response facilities for SGRB studies.
Throughout the paper, we will use the standard cosmological parameters,
$H_0 =70\ $km s$^{-1}$ Mpc$^{-1}$, $\Omega_m =0.27$, and $\Omega_{\Lambda}=0.73$.

\section{Observations and Data Reduction}
\label{sec:data}
On June 3, 2013, at $T_0$=15:49:14 UT, the \swift\ satellite \citep{Gehrels:2004fj} triggered on \grb\ \citep{melandri:2013aa}.
The on-board Burst Alert Telescope \citep[BAT;][]{Barthelmy:2005lr} detected 
a single bright peak with a duration of 0.4 seconds, placing this event unambiguously in
the short-GRB category \citep{norris:2013aa}.  After slewing to the source location, the 
\xray\ Telescope \citep[XRT; ][]{Burrows:2005fk},
began observing 59\,s after the trigger, and detected a fading \xray\ counterpart at
 $\alpha =11^{\rm h}$28$^{\rm m}$48$^{\rm s}$.16, $\delta =+17\degr$04$'$18\arcsec.8
\citep[with an uncertainty of 2.7\arcsec;][]{Evans:2013aa}. No optical counterpart was found in the UV/Optical Telescope
\cite[UVOT; ][]{Roming:2005qy}  prompt data \citep{melandri:2013aa}, 
while a counterpart was detected when more data became available \citep{de-Pasquale:2013cr}.

At $T_0+5.8$ hours, using the William Herschel Telescope, \cite{Levan:2013aa}
identified a point-like source
inside the XRT error circle, which they determined to be the optical counterpart of the short \grb. 
This position lies in the outskirts of a galaxy present in the archival Sloan Digital Sky Survey DR9 \citep{Ahn:2012ly}.
Other subsequent follow-up observations were performed on \grb\
yielding a spectroscopic redshift for the afterglow and/or the host galaxy \citep{thoene:2013aa,xu:2013aa,sanchez:2013aa,Foley:2013aa, cucchiara:2013aa}. The afterglow was also later detected at radio wavelengths \citep{fong:2013aa}.
Unfortunately, no observations to date have been providing 
detections of fine-structure lines, which are connected to the GRB radiation itself (leaving still some uncertainty on the actual GRB-host association).

\begin{figure*}[t!]
\epsscale{1.0}
\includegraphics[scale=0.50,angle=0]{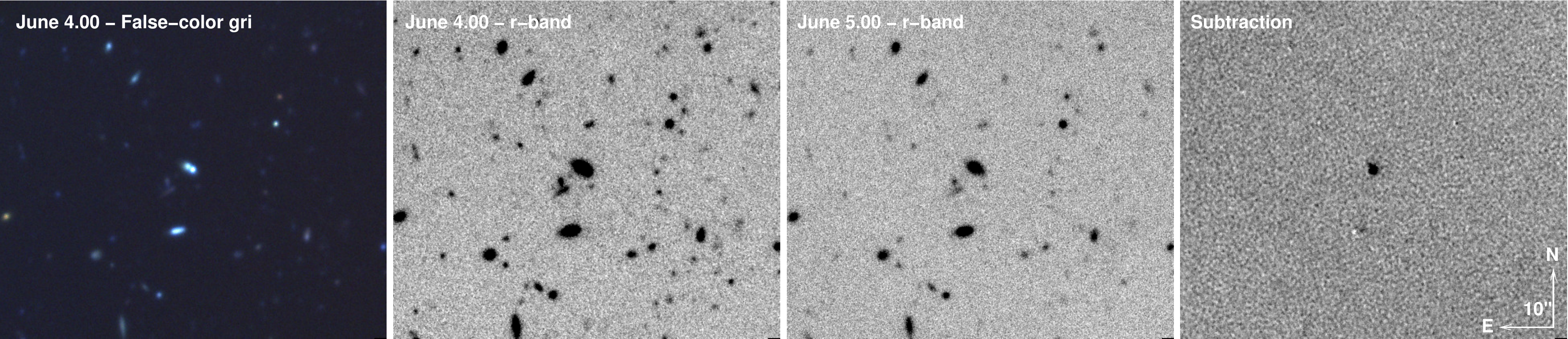}
\caption{From left to right: composite false-color image of \grb\ and
  associated galaxy obtained the night of the discovery with
  Gemini/GMOS; \rp\ band coadded image obtained the first night; \rp\
  band coadded image obtained the second night; host galaxy subtracted
  image of the GRB.} \label{fig:finder}  
\end{figure*}

Using our ToO program (GS-2013A-Q-31; PI Cucchiara) 
we performed a series of photometric observations with the GMOS camera \citep{Hook:2004fj} on Gemini South in the \gp, \rp, and \ip\ filters for a total of 8x180\,s exposures per band from $T=T_0+7.19$\,h to $T=T_0+9.1$\,h. 
The data were analyzed using the standard {\tt GEMINI/GMOS} data analysis packages within the IRAF\footnote{IRAF is distributed by the National Optical 
Astronomy Observatory, which is operated by the Association for Research 
in Astronomy, Inc., under cooperative agreement with the National Science 
Foundation.} environment.
The afterglow is detected at a projected distance of $\approx 0.92\pm 0.10 \arcsec$ 
from the center of a bright, neighboring galaxy (see the false-color image in Fig. \ref{fig:finder}).
At the galaxy's redshift (see next section), this corresponds to
4.8 ($\pm 0.5$)\,kpc in projected distance.

\begin{figure*}[t!]
\epsscale{1.0}
\includegraphics[scale=0.60,angle=0]{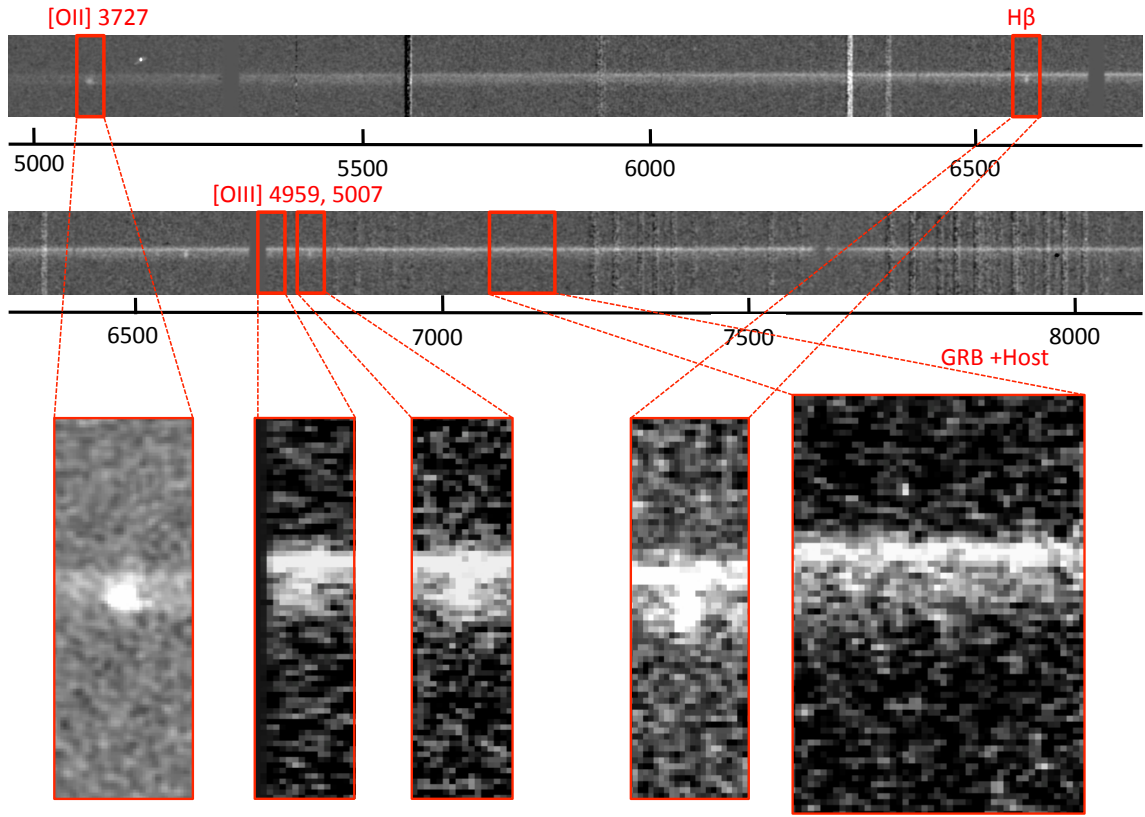}
\caption{Two-dimensional image of the spectrum of \grb\ covering the
5000-8000\AA\ range. The bottom panels show a zoom-in view of some of the most
prominent nebular lines (first four panels) and a region free of any emission line
(last panel) to emphasize the presence of the strong GRB afterglow
emission superimposed on the faint host galaxy.}  
\label{fig:twod}
\end{figure*}

Subsequently, we obtained a spectroscopic sequence with the same instrument:
we obtained 2x900\,s spectra, using the R400 grism with the 1\arcsec\ slit (resolution of about 5.5 \AA)  centered at 6000 \AA, 
covering wavelengths $3900-8100$\,\AA.
We reduced the spectroscopic data with standard techniques, 
performing flat-fielding, wavelength calibration with CuAr lamp spectra, 
 and cosmic ray rejection using the {\tt lacos\_spec} package \citep{dok01}. 
A sky region close in the spatial direction, but unaffected by the spectral trace, was used for sky subtraction.
The two-dimensional spectra were then coadded. 
Figure \ref{fig:twod} presents the processed data which reveal
the spectrum of the extended galaxy exhibiting a faint continuum and a
series of emission lines. Clearly visible superimposed on the galaxy light is an unresolved (spatially) trace that coincides with the expected position of the afterglow based on our imaging data.  

Using the {\tt IRAF/APALL} routine, we extracted a spectrum corresponding 
to the entire detected trace, 
therefore including both the galaxy and the GRB afterglow signal.  We then extracted
a second spectrum with the aperture restricted to the spatial location
of the GRB afterglow. Variance spectra were extracted in both cases
evaluating sky contribution in two regions unaffected by the galaxy or the GRB light
(plotted in grey in Fig.~\ref{fig:comparis} and with dash lines in Fig. ~\ref{fig:twod}).

Flux calibration was performed using an observation
of the spectrophotometric standard star Feige~110 taken with the same
instrument configuration, although no correction 
has been made for slit losses.
An air-to-vaccum correction was applied to the 1D wavelength solution. 
Using the measured optical brightness of the GRB+host we also corrected for a small slit-loss ($\lesssim 10\%$).

On the following day, starting at $T_0+ 1.3$ days after the burst, we observed again 
the field with Gemini South obtaining $3\times180$\,s 
imaging exposures with the GMOS camera in the \gp, \rp, \ip\ bands and a single 900\,s  
spectrum with the same configuration
used the first night. The afterglow had faded considerably in comparison with the host 
galaxy. The reduced spectrum revealed no
trace from the GRB, but only a faint continuum from the host, with identical emission 
lines superimposed.

We also obtained an optical spectrum of the host galaxy of \grb\ on 2013 June 6 UT with the Deep Imaging Multi-Object Spectrograph \citep[DEIMOS;][]{Faber:2003tg} mounted on the 10 m Keck II telescope. The instrument was configured with the
600 lines mm$^{-1}$ grating, providing spectral coverage over the region $\lambda = 4500$--9500\,\AA\ with a spectral resolution of 3.5\,\AA.  The spectra were optimally extracted \citep{Horne:1986dq}, and the rectification and sky subtraction were performed following the procedure described by \cite{Kelson:2003hc}. The slit was oriented at an angle such to include the host nucleus and the GRB location. Flux calibration was performed relative to the spectrophotometric standard star BD+262606.

Finally, on June 16 UT, we imaged the field with the Gemini-South telescope in \gp, \rp,\ip, and \zp\
bands to estimate the galaxy contribution at a time when the afterglow was expected to have faded
well beyond our detectability.

\section{Spectral Analysis}
\label{sec:analys}
\begin{figure*}[tl!]
\epsscale{1.0}
\includegraphics[scale=0.7,angle=0]{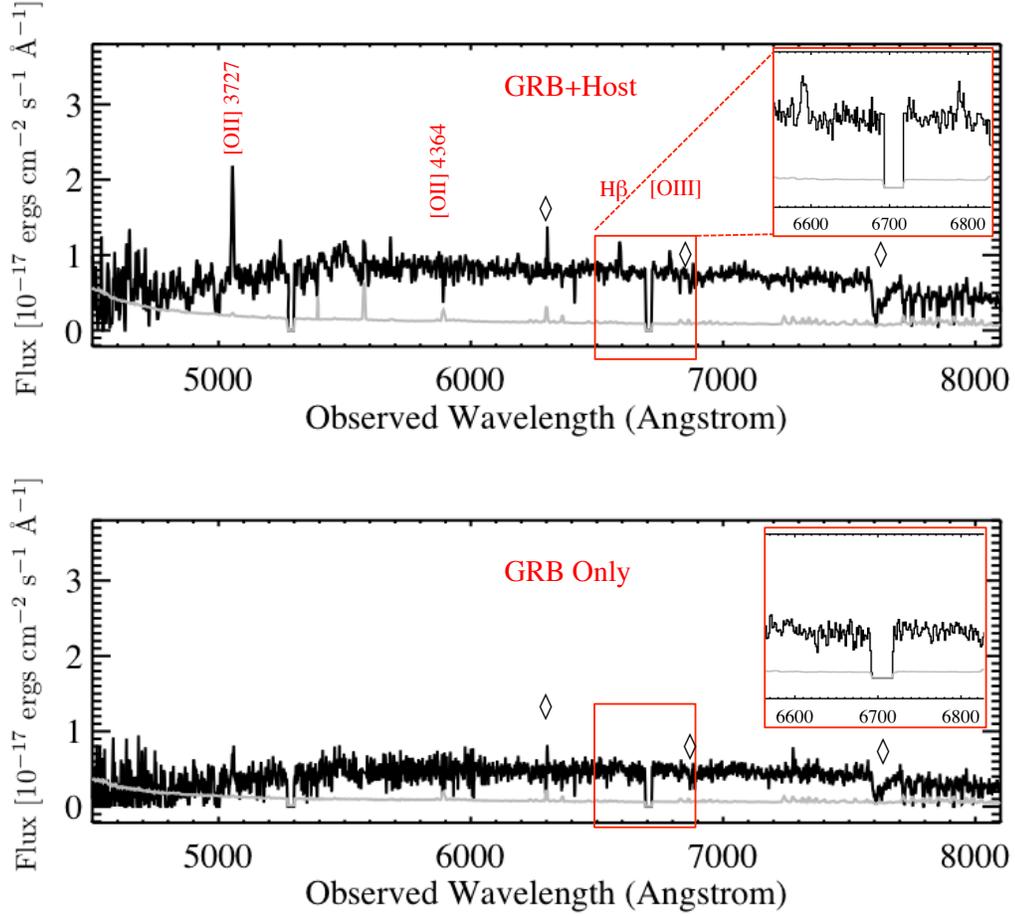}
\caption{{\emph Top:} GMOS spectrum of
the GRB and the host galaxy using the data obtained the first night (black are the data, gray the variance spectrum,
and diamonds indicate the strongest atmospheric telluric lines). 
Inset shows a zoom-in section at the location of the \Hb\ and \OIII4959,5007 lines. 
{\emph Bottom: same as at the $Top$, but
extracting solely the spectrum at the GRB location. In the inset we
show a zoom-in section: no emission lines are detected at those locations while
very little emission is present at the location of \OIII3727, indicating 
the GRB explosion location is not a region of especially
active star-formation.}}  
\label{fig:comparis}
\end{figure*}

Figure~\ref{fig:comparis} presents extractions associated with the combined afterglow and galaxy and for an aperture restricted to the afterglow location.
In the combined spectrum, we identify a series of nebular emission
lines (e.g. \OII$\lambda3727$, \OIII$\lambda \lambda$4959,5007, 
and \Hb\ ), at a common redshift $z= 0.3568 \pm 0.0005$.
We associate these lines to \ion{H}{2} regions near the center of the galaxy, 
spatially offset from the afterglow location (Fig.~\ref{fig:twod}).
The only significant absorption features present in the spectrum are
\CaII\,H+K features, despite the fact that the flux and S/N are lower (see
 Sec. \ref{sec:ot})

The narrow aperture centered on the afterglow location shows a largely featureless continuum (lower panel of Figure~\ref{fig:comparis}).  
At wavelengths $\lambda \gtrsim 6000$\,\AA\ the two-dimensional spectrum and our
imaging indicate the afterglow dominates the flux.  We measure a
spectral slope in the optical of $\nu f_\nu \propto
\nu^{-\beta_o}$ with $\beta_o = 0.62 \pm 0.17$.    

Examining the GRB afterglow spectrum at the wavelengths of the galaxy's nebular
lines we detect little or no  emission (see zoom-in regions in Fig. \ref{fig:comparis}).  
Upper limits on the line fluxes are given in Table~\ref{tab:host}.

\begin{figure*}[t!]
\epsscale{1.0}
\includegraphics[scale=0.7,angle=0]{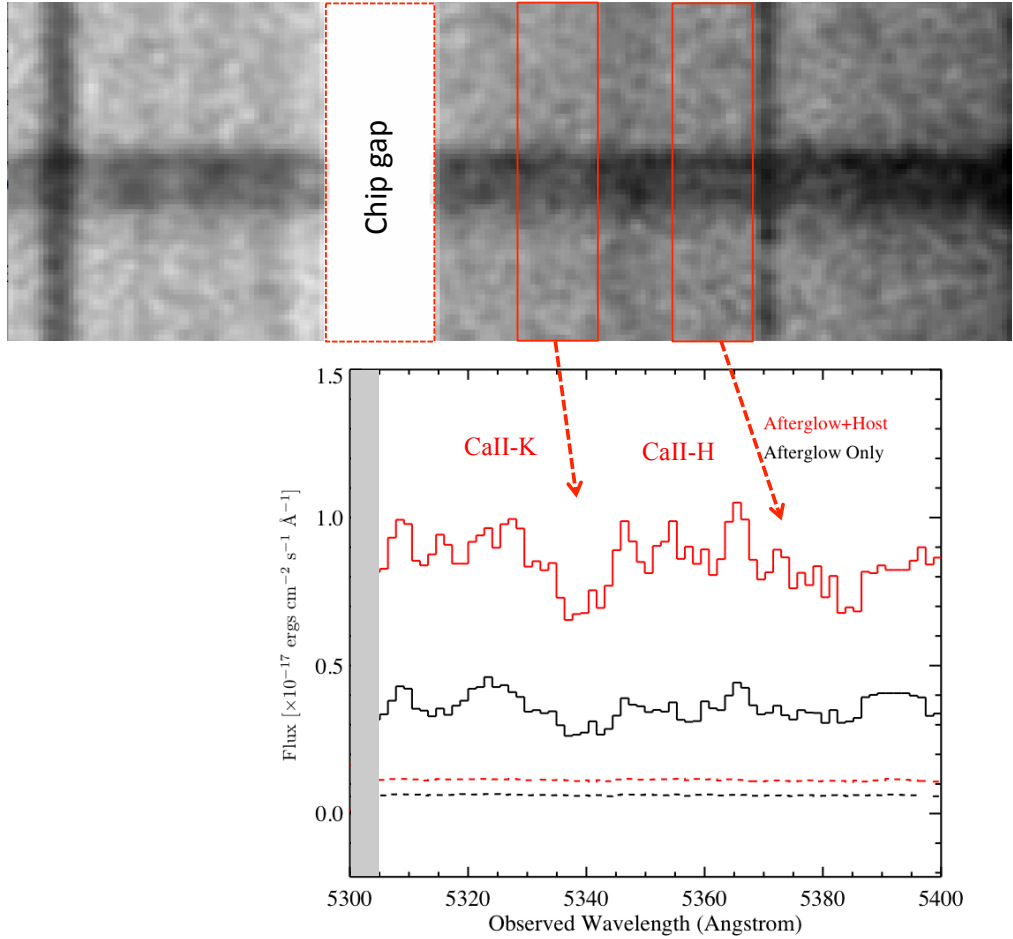}
\caption{Section of the two-dimensional spectrum centered
on the CaII H\&K absorption features, re-binned in order to enhance the
color contrast (top panel). The afterglow and host traces are well visible and at the position
of the CaII lines (in particular the K line at $\lambda_{rest}=3933.7$\,\AA) there is a 
decrement in the continuum flux. In the bottom panel we present the same region for the extracted one-dimensional spectra
(black and red for the Afterglow only or the Afterglow+Host galaxy traces respectively, while in dash we plot the sigma spectra). 
Also in this case, despite the low S/N,
the absorption features are evident and allow, in combination with the emission lines (see Fig.\ref{fig:twod}) to secure the redshift of the GRB at $z=0.3568\pm 0.0005$.}
\label{fig:HK}
\end{figure*}

Figure~\ref{fig:HK} shows a section of the two-dimensional spectrum centered on  
the location of the \CaII\ doublet (dashed lines represent the variance spectrum).
In the outset we highlight this region in the corresponding extracted spectra for the afterglow only 
and the afterglow+host: at the location of the redshifted \CaII\ H\&K we see 
a decrement in flux in the two-dimensional spectrum and in both extracted spectra.
Using the variance spectra we can determine the pixel-by-pixel flux error. 
At the expected location we detect a redshifted \CaII\ K-line at $\sim4\sigma$ significance level 
in the afterglow+galaxy spectrum
($2.5\sigma$ in the afterglow only), while
the \CaII\ H-line only at $2\sigma$ (and $\lesssim 1\sigma)$. 
This result, in combination with the emission lines detected, places strong constraints on the GRB redshift  
($z\gtrsim 0.3568$) and suggests a likely association of the GRB with the galaxy.

\section{Results}
\label{sec:results}

\subsection{Properties of the GRB and its Afterglow}

Our second and third imaging epochs do not show any clear sign of the afterglow either in the single exposures nor in coadded ones.
Assuming that the afterglow completely faded below our detection limit at $T_0 +1.3$ days
we subtract from our first epoch of imaging this ``reference'' to measure the afterglow 
flux at the time of our initial observation.  Using the HOTPANTS\footnote{See http://
www.astro.washington.edu/users/becker/hotpants.html.} code and nearby point sources 
from SDSS for calibration, we measure the following, extinction corrected, AB magnitudes: $\gp=22.09\ (\pm0.04), \rp=21.52\ (\pm0.05)$, and $\ip=21.18\ (\pm 0.11)$.
We derive, at $T_0+0.35$ days, an afterglow brightness of $\rp=21.52 \pm0.05$, after correcting for Galactic
extinction ($E$(B-V) =0.02): this value is similar to other
optically-detected short GRBs observed around the same epoch \citep{Nicuesa-Guelbenzu:2012dq,Kann:2012ve,Kann:2011qf,Berger:2010cr}. 
Moreover, using the data from June 16 as 
a ``reference'' does not change these results, while a similar  procedure operated between our second and third epochs, allows
us to place stringent upper limits on the afterglow brightness at $T_0+1.3$ days ($\rp > 25.2$ and $\ip > 24.5.$), confirming our assumption that indeed the optical afterglow was undetectable by Gemini. Interestingly, almost at similar time of our latest Gemini observation, a near-IR
counterpart was still visible in the $HST$ data as reported by \cite{Berger:2013ij} and \cite{Tanvir:2013hc}.

The peak of the galaxy emission and the centroid of the afterglow
emission derived from the subtracted image (see last panel on the
right in Fig. \ref{fig:finder}) are separated by $\delta r =
0.92\arcsec \pm 0.10$, corresponding to a projected distance of 4.8\,kpc ($\pm 0.5$)
at \zgrb.
At this redshift the probability of a random association between the galaxy and the GRB at such distance $\delta r$ is very small \citep[$P(< \delta r) < 10^{-3}$; see also][]{Bloom:2002ys}.

Based on our $g', r', i'$ photometry we derive an 
optical spectral index $\beta_o = 0.54\pm 0.12$ (assuming $\nu
F_{o,\nu} \propto \nu^{- \beta_o}$) consistent with 
analysis of our optical spectrum and similar to other SGRBs \citep{Nicuesa-Guelbenzu:2012dq}.

\subsection{Properties of the Host Galaxy} 
\begin{figure*}[t!]
\epsscale{1.0}
\includegraphics[scale=0.70,angle=0]{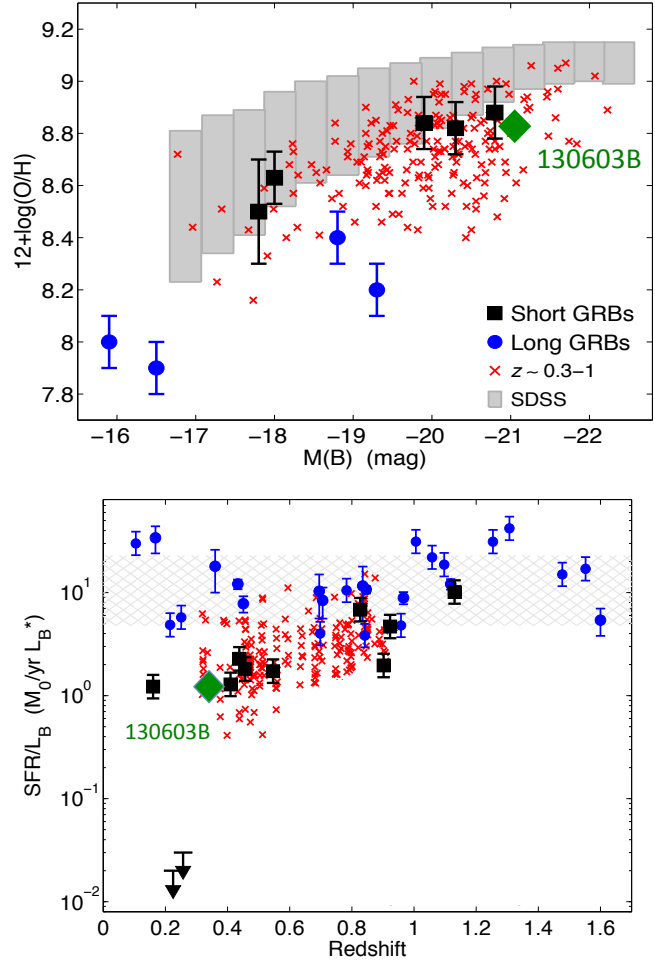}
\caption{Adapted from Berger et al. 2010. {\emph Top:} Metallicity as function of host galaxy absolute B-magnitude of long (blue points) and short (black squares). Crosses are field galaxies from  \cite{Kobulnicky:2004bs}, while grey bars are SDSS galaxy at $z\sim0.1$. \grb\ is indicated with a green diamond and clearly shows similar property of the short GRBs host population. $Bottom:$ Specific star formation rate vs. redshift for long (blue), short (black) GRB hosts as well as field galaxies (red crosses). Again the green diamond marks the short \grb. It is evident that the host of \grb\ is very similar to other SGRB hosts, though the properties of the GRB explosion site could be very different from its host center.} 
\label{fig:sfr}
\end{figure*}

At \zgrb\ the \rp\ band magnitude samples
the rest-frame B-band luminosity of the host. 
Therefore, using our second epoch observations 
we derive  \rp=20.76 $\pm 0.06$\,mag for the GRB host and 
we estimate a rest-frame, $k$-corrected, absolute $B$-band magnitude (AB) of
$M_B= -20.96$. Despite the brightness of this galaxy \citep[$\sim L^*$;][]{Zucca:2009pi}, the derived luminosity is not unusual among the 
short-GRB hosts population \citep[Figure \ref{fig:sfr}, ][]{Berger:2009nx}. 
Using the late-time multiband photometry and the IDL package \emph{kcorrect} 
\citep[][]{Blanton:2007tg} in a similar fashion to \cite{Werk:2012kl} we estimate the mass 
of the host galaxy as $\mathcal{M}\approx 5.0 \times 10^9 \,M_{\odot}$.

The GMOS spectra from the first two nights and the DEIMOS data
cover key nebular emission lines.
We measured the fluxes of these lines using the latter data, but for the ones
also detected in the GMOS spectra we  obtain similar results
(Table~\ref{tab:host}). 
We  apply standard emission-line analysis to
derive intrinsic galaxy properties. 
We have estimated the optical extinction using the Balmer lines decrement
and assuming case-B recombination \cite{Calzetti:1994oq,Kennicutt:1998nx}:
we derived E(B-V) = 0.43 ($A_V=1.3$ mag, assuming a Milky-Way extinction curve).
Using the calibration of \citep{ken98}, the [OII] line luminosity
gives a SFR([OII])\ $ = 1.7 \ M_\odot$ yr$^{-1}$. 
 Similarly, from the \Ha\ luminosity we derive 
 SFR(\Ha) $=$  \sfr. 
Together with the $B$-band lumiosity, we derive a specific SFR: 
sSFR=SFR$/L^*_B\gtrsim2.1 M_{\odot}$ yr$^{-1}\ L^{*-1}$. 
As in other many short GRB hosts, this value is consistent 
with a star-forming galaxy (see Fig.\ref{fig:sfr}, Berger 2009).

Also, collisionally-excited oxygen
and the \Ha\ and \Hb\ Balmer series recombination lines provide an 
estimate of the gas-phase oxygen abundances of the
host galaxy.  We adopt the 
$R_{23}=(F_{\OII \lambda3727}+F_{\OIII \lambda
  \lambda4959,5007}$)/$F_{\Hb}$ 
metallicity indicator \citep{Kobulnicky:2004bs,Pagel:1979ij}, which 
depends on both the metallicity and the ionization state of the gas.
To help disentangle a degeneracy in the values,
we use the $O_{32}$ indicator 
($O_{32}=F_{\OIII \lambda \lambda4959,5007}$/$F_{\OII \lambda3727}$).

However, our measured fluxes still allow for two solutions:
$\approx 8.7$ for the upper $R_{23}$ branch and $\approx 7.8-8.3$ for
the lower branch. The typical error in this measurement, due to
systematic uncertainty in the calibration of these metallicity
relations is typically $\sim0.2$ dex. No field galaxy at $z \sim 0.4$ 
with a similar brightness to \grb\ exhibits metallicities consistent with
the lower-branch of our $R_{23}$ analysis.
Furthermore, based on the detection of \Ha\ and \NII$\,\lambda{6583}$ lines
in the DEIMOS spectrum we place
a lower limit on the metallicity of 12 + log(O/H) $\gtrsim 8.5$.
So we can conclude the galaxy has approximately solar 
metallicity \citep{Asplund:2005fk}.

\subsection{The Afterglow Spectrum of \grb}
\label{sec:ot}

GRB\,130603B marks one of the best cases in which, 
for a short GRB, an afterglow-dominated spectrum has been obtained.
Overall, the extracted GRB afterglow spectrum presents a smooth, almost featureless 
continuum (Figure \ref{fig:comparis} lower panels).
Despite the absence of strong spectral features, 
we can still constrain the GRB redshift from these data alone: 
1) the lack of a \lya\ forest requires that
the GRB exploded at $z_{\rm GRB}<2.9$; 
2) the appearance of weak \CaII\, H+K absorption lines (see
Figure \ref{fig:HK})
as also seen by \citet[][]{thoene:2013aa} and \citet{sanchez:2013aa}, gives a
strong indication that the GRB redshift is at least a the the emission line redshift
($z_{\rm GRB} \gtrsim 0.3586$); 
3) if one assumes the afterglow spectrum will exhibit significant
($>0.3$\,\AA) \MgII\ absorption from the surrounding interstellar or
circumgalactic medium then we set an upper limit
$z_{\rm GRB} <0.78$ based on its non-detection. 
All of these constraints are consistent with \grb\ having occurred within 
the coincident galaxy.
Ultimately, an indisputable measurement of $z_{\rm GRB}$ from an
afterglow spectrum may require the detection of fine-structure
transitions excited by the GRB itself \citep{pcb+06} and/or the
observed termination of the \lya\ forest. Unfortunately, our wavelength coverage 
and the small number of this kind of features that would have been redshifted 
in our observed window (and therefore detected) prevent us to obtain such 
a secure measurement.
Furthermore, analysis of these transitions would reveal physical
properties of the progenitor environment (e.g.\ metallicity, size)
and may offer new insight into the progenitors of SGRBs.

Associating \grb\ to the neighboring galaxy, we may place an upper
limit to the star-formation rate at the position of the event.
No strong emission lines are identified, in contrast
to the putative host spectrum: the striking difference can be seen at
the location of the \Hb\ and \OIII$\lambda \lambda$4959,5007 lines
(insets in Fig.  \ref{fig:comparis}).  
This indicates that \grb, unlike the majority of
long-duration GRBs, exploded in 
a region of very minimal, if not negligible, star-formation.
Based on the integrated flux at the location
of the \Hb\ line, we place an upper limit 
SFR(\Hb) $\lesssim 0.4\ M_\odot$ yr$^{-1}$.
This provides further support for models unassociated with massive star formation.

\section{Summary and Conclusions}
\label{sec:conclusions}
We present our Gemini rapid follow-up campaign on the afterglow 
of the short \grb.
Despite the intrinsic faintness of these events, the low-energy 
emission of \grb, in particular in the optical afterglow, was still detectable several hours after the explosion.
We triggered our approved ToO program at the Gemini South telescope and obtained a series
of images of the GRB field as well as spectroscopic observations with the GMOS camera starting 8 hours after the burst.
We repeated similar observations the following night.

Based on our absorption
lines analysis and the very small probability of a random association with such a bright galaxy  ($P(\leq \delta r)$=0.00064) we conclude that
this is the host of \grb.
Our optical images provide firm evidence that this event occurred in the outskirt of a star-forming galaxy ($M_B=-20.96$),
around 4.8\,kpc from its center (at the GRB redshift of $z_{grb}=0.3568$). 
While this offset is not unusual for long duration GRBs \citep[e.g., ][]{Bloom:2002ys}, the lack of blue light at the location is very unlike long-duration GRB locations \citep{Fruchter:2006lh}. Deep $HST$ observations also confirm this findings \citep{Berger:2013ij,Tanvir:2013hc}.

The early-time spectra present 
a bright afterglow continuum superimposed upon a 
fainter galaxy. This represents one of the few cases where an afterglow-dominated
spectrum for a short GRB has been recorded. 
The host  is an $\sim L^*$ galaxy, and the spectrum shows strong
nebular emission lines as well as recombination features. We measure a dust-corrected
SFR $=$\sfr\, and  solar metallicity. 

Nevertheless, we were able to extract the GRB spectrum and constrain the properties
of the host galaxy at its location: the GRB spectrum present a smooth continuum showing almost no sign
of emission lines throughout the wavelength coverage. 
Also, at the same redshift we identify \CaII\ absorption
lines, which place a strong constraint to the GRB redshift, despite the lack of fine-structure transitions.
Therefore, at the GRB location, the star-formation activity is almost negligible 
(SFR(\Hb) $\lesssim 0.4\ M_\odot$ yr$^{-1}$). 

This last result, once more, emphasizes the importance of rapid
follow-up observations with large aperture facilities 
in order to firmly identify the faint afterglows of these short-lived
events. Only with a larger sample of data   
similar to those presented in this work we will be able not only to identify
a larger number of short GRBs and measure their redshifts, but also to 
characterize, via absorption spectroscopy, their explosion site environment
and move forward in our understanding of the nature of their
progenitors.

\acknowledgements
AC thanks the anonymous referee for the valuable comments and suggestions which have helped significantly improving 
the manuscript.
Gemini results are  based on observations obtained at the Gemini Observatory, which is operated by the 
    Association of Universities for Research in Astronomy, Inc., under a cooperative agreement 
    with the NSF on behalf of the Gemini partnership: the National Science Foundation 
    (United States), the National Research Council (Canada), CONICYT (Chile), the Australian 
    Research Council (Australia), Minist\'{e}rio da Ci\^{e}ncia, Tecnologia e Inova\c{c}\~{a}o 
    (Brazil) and Ministerio de Ciencia, Tecnolog\'{i}a e Innovaci\'{o}n Productiva (Argentina).
Partial support for this work was provided by NASA through Hubble Fellowship
grant HST-HF-51296.01-A awarded by the Space Telescope Science
Institute, which is operated by the Association of Universities for
Research in Astronomy, Inc., for NASA, under contract NAS 5-26555.

\begin{deluxetable}{ll}
\tablewidth{0in}
\tablecaption{Properties of the Host and Afterglow \label{tab:host}}
\tabletypesize{\footnotesize} 
\tablehead{
\colhead{} & \colhead{Value} 
} 
\startdata
$\alpha_{Host}$ &11:28:48.227\\
$\delta_{Host}$  &+17:04:18.42\\
$\alpha_{OT}$ &11:28:48.168\\
$\delta_{OT}$  &+17:04:18.06\\
$z$ &$0.3568 \pm 0.0005$\\
Separation$^a$ ($\delta r$) &0.85\arcsec\ W, 0.36\arcsec\ S \\
P($\leq \delta r$) & 0.00064$^b$\\
$F_{\OII \lambda 3727}$&$28.2 \,(\pm 0.4)^c$\\
$F^d_{GRB, \OII \lambda 3727}$&$\leq5.8^c$\\
$F_{\Hb}$&$12.6\, (\pm 0.3)^c$\\
$F^d_{GRB, \Hb}$&$\leq1.44^c$\\
$F_{\OIII \lambda 4960}$&$1.89\, (\pm 0.12)^c$\\
$F^d_{GRB,\OIII \lambda 4960}$&$\leq1.45^c$\\
$F_{\OIII \lambda 5008}$&$8.25\, (\pm 0.23)^c$\\
$F^d_{GRB,\OIII \lambda 5008}$&$\leq2.31^c$\\
$F_{\Ha}$&$54.0\, (\pm 0.2)^c$\\
$F_{\NII \lambda 6584}$&$13.8 \,(\pm 0.3)^c$\\
$M_{Host, \rm B}$ &-20.96 $(\pm 0.07)^e$\\
SFR$_{Host}$ & \sfr \\
12 + log(O/H) &8.7 $\pm 0.2$\\
$\mathcal{M}$ & $5 \times 10^{9}$ M$_{\odot}$\\
$A_V$ & 1.3 mag\\
\enddata
\tablenotetext{a}{0.92\arcsec\ ($\pm 0.10$) projected or 4.8 $\pm 0.5$\,kpc}
\tablenotetext{b}{Probability to find such a host galaxy at projected distance
$\leq \delta r$. }
\tablenotetext{c}{in units of $10^{-17}$ erg cm$^{-2}$ s$^{-1}$}
\tablenotetext{d}{Upper limit on the galaxy line-flux measured from
 the GRB aftergow spectrum.}

\tablenotetext{e}{Rest-frame B magnitude derived from observed \rp.}

\end{deluxetable}



%

%
%

\clearpage

\clearpage



\end{document}